\documentclass[prl,aps,amssymb,twocolumn]{revtex4-1}
\usepackage{amsmath}
\usepackage{amssymb}
\usepackage{amsthm}
\usepackage{amsfonts}
\usepackage{listings}
\usepackage{tabularx}
\usepackage{makecell}
\usepackage{soul, color,xcolor}
\usepackage{enumerate}
\usepackage{latexsym}
\usepackage{color}
\usepackage{bm}
\usepackage{hyperref}
\hypersetup{
 pdfnewwindow=true, colorlinks=true,
 linkcolor=blue, anchorcolor=blue,
 citecolor=blue, filecolor=blue,
 menucolor=blue, urlcolor=blue}

\usepackage{psfrag}

\usepackage{bm}
\usepackage{graphicx}
\usepackage{subfigure}

\soulregister{\cite}7
\soulregister{\ref}7
\RequirePackage[normalem]{ulem} 
\RequirePackage{color}\definecolor{RED}{rgb}{1,0,0}\definecolor{BLUE}{rgb}{0,0,1} 

\newcommand{\bk}{{\bf k}}

\input{epsf}

\newcommand{\nc}{\newcommand}
\nc{\fig}[1]{Fig.\ref{#1}}
\nc{\webirvsp}{\href{https://github.com/zjwang11/irvsp}{\texttt{IRVSP}} }
\nc{\webirtb}{\href{https://github.com/zjwang11/irvsp}{\texttt{ir2tb}} }
\nc{\webirpw}{\href{https://github.com/zjwang11/ir2pw}{\texttt{IR2PH}} }
\nc{\webvaspkp}{\href{https://github.com/zjwang11/VASP2KP}{\texttt{VASP2KP}} }
\nc{\webchecktopmat}{\href{https://www.cryst.ehu.es/cryst/checktopologicalmagmat}{\texttt{Check Topological Mat}}}
\nc{\webposbr}{\href{https://github.com/zjwang11/UnconvMat/blob/master/src_pos2aBR.tar.gz}{\texttt{POS2ABR}} }
\nc{\webUnconvMat}{\href{http://tm.iphy.ac.cn/UnconvMat.html}{\texttt{UnconvMat}} }
\nc{\online}{\href{http://tm.iphy.ac.cn/UnconvMat.html}{online}}
\nc{\arxivjia}{\href{https://arxiv.org/abs/2403.07637}{arXiv:2403.07637(2024)}}

\begin{document}

\tolerance 10000

\newcommand{\vk}{{\bf k}}

\draft

\title{Axion insulator, Weyl points, quantum anomalous Hall effect \\and magnetic topological phase transition in Eu$_3$In$_2$As$_4$}


\author{Jingyu Yao}
\thanks{These authors contributed equally to this work.}
\affiliation{Beijing National Laboratory for Condensed Matter Physics,
and Institute of Physics, Chinese Academy of Sciences, Beijing 100190, China}
\affiliation{University of Chinese Academy of Sciences, Beijing 100049, China}

\author{Ruihan Zhang}
\thanks{These authors contributed equally to this work.}
\affiliation{Beijing National Laboratory for Condensed Matter Physics,
and Institute of Physics, Chinese Academy of Sciences, Beijing 100190, China}
\affiliation{University of Chinese Academy of Sciences, Beijing 100049, China}

\author{Sheng Zhang}
\affiliation{Beijing National Laboratory for Condensed Matter Physics,
and Institute of Physics, Chinese Academy of Sciences, Beijing 100190, China}
\affiliation{University of Chinese Academy of Sciences, Beijing 100049, China}
\author{Haohao Sheng}
\affiliation{Beijing National Laboratory for Condensed Matter Physics,
and Institute of Physics, Chinese Academy of Sciences, Beijing 100190, China}
\affiliation{University of Chinese Academy of Sciences, Beijing 100049, China}

\author{Youguo Shi}
\affiliation{Beijing National Laboratory for Condensed Matter Physics,
and Institute of Physics, Chinese Academy of Sciences, Beijing 100190, China}
\affiliation{University of Chinese Academy of Sciences, Beijing 100049, China}

\author{Zhong Fang}
\affiliation{Beijing National Laboratory for Condensed Matter Physics,
and Institute of Physics, Chinese Academy of Sciences, Beijing 100190, China}
\affiliation{University of Chinese Academy of Sciences, Beijing 100049, China}

\author{Hongming Weng}
\affiliation{Beijing National Laboratory for Condensed Matter Physics,
and Institute of Physics, Chinese Academy of Sciences, Beijing 100190, China}
\affiliation{University of Chinese Academy of Sciences, Beijing 100049, China}

\author{Zhijun Wang}
\email{wzj@iphy.ac.cn}
\affiliation{Beijing National Laboratory for Condensed Matter Physics,
and Institute of Physics, Chinese Academy of Sciences, Beijing 100190, China}
\affiliation{University of Chinese Academy of Sciences, Beijing 100049, China}

\begin{abstract}
The magnetic topological phases attract much interest, such as the axion insulator, higher-order topology, Weyl semimetals, and the quantum anomalous Hall effect (QAHE).
Here, we predict that the axion insulator phase, magnetic Weyl points, and QAHE can be achieved in Eu$_3$In$_2$As$_4$. Recently, the single-crystal Eu$_3$In$_2$As$_4$ has been successfully synthesized, which exhibits an antiferromagnetic (AFM) ground state. Our first-principles calculations show that it lies on the phase boundary between multiple magnetic topological phases, and the magnetic anisotropy is weak, with an energy difference less than 1 meV. In the AFM state, it can be tuned to an axion insulator by tensile strain. The quantized axion angle $\theta = \pi$ and the magnetic higher-order topology are characterized by the parity index $Z_4=2$. By applying an external magnetic field, the induced ferromagnetic (FM) state becomes an ideal magnetic topological semimetal with a single pair of Weyl points or a nodal ring. The QAHE can be achieved in FM  multilayer films of Eu$_3$In$_2$As$_4$ on a magnetic insulating substrate.

\end{abstract}

\maketitle
\paragraph*{Introduction.---}In recent decades, topological phases of matter have gained considerable interest in condensed matter physics and materials science. Notably, the intriguing interaction between magnetism and topology can lead to a variety of exotic topological quantum phenomena, such as the quantum anomalous Hall effect~\cite{qah-PhysRevLett.61.2015,qah-doi:10.1126/science.1187485,qah-doi:10.1126/science.1234414}, a single pair of Weyl points (WPs)~\cite{Weyl-one-BAB,PhysRevLett.124.239901,PhysRevB.83.205101}, and axion electrodynamics~\cite{axion-PhysRevLett.58.1799,axion-PhysRevB.78.195424,axion-PhysRevLett.102.146805,axion-nphys1534,axion-JAP10.1063/5.0038804}.
In condensed matter physics, the axion field appears within the field theory framework of the topological magnetoelectric effect~\cite{axion-PhysRevLett.58.1799,axion-PhysRevB.78.195424}.  Its effective action can be expressed in the following form~\cite{axion-PhysRevB.78.195424},
$S_\theta=\cfrac{\theta e^2}{4\pi^2}\int dt d^3 \boldsymbol{x} \boldsymbol{E}\cdot \boldsymbol{B}$,
where $E$ and $B$ represent the electromagnetic fields, and the coefficient $\theta$ is the axion angle with a period of $2\pi$. 
A magnetic topological crystalline insulator, with inversion symmetry or any other antiunitary symmetry operation  (e.g., $T\tau$, $TC_n$) that reverses an odd number of space–time coordinates, can quantize the axion angle $\theta$ to values of 0 or $\pi$~\cite{PhysRevB.91.161105,Rosa-npjQM2020}.
In inversion-invariant magnetic systems, the magnetic parity index is well defined~\cite{EuInAs122-PhysRevLett.122.256402,Zhang2020,axion-hinge-Gu2021}, 
\begin{equation}
        Z_4=\sum_{\boldsymbol{k}=1}^{8}        \frac{n^{+}_{\bm{k}}-n^{-}_{\bm{k}}}{2} \mod{4}, 
\end{equation}
where $n^{\pm}_{\bm{k}}$ represent the number of occupied states with $+/-$ parity at one of the eight inversion-invariant momenta $\boldsymbol{k}$. $Z_4=2$ indicates an axion insulator (AI) and higher-order topological phases.
The AI state is expected to emerge in magnetic heterostructures of topological insulators (TIs)~\cite{axion-heter-nmat4855,axion-heter-PhysRevLett.120.056801,axion-heter-sciadv.aao1669} or in TIs with external magnetic fields or magnetic doping~\cite{axion-exter-science.aaf5541,axion-hinge-ycm,PhysRevB.105.195105}.
However, intrinsic AIs are quite rare, with notable examples including MnBi$_2$Te$_4$~\cite{MnBiTe-NM10.1038,MnBiTe-PhysRevLett.122.206401,MnBiTe-sciadv.aaw5685,Otrokov2019,PhysRevLett.122.107202}, Mn$_2$Bi$_2$Te$_5$~\cite{PhysRevB.102.121107}, and EuIn$_2$As$_2$~\cite{EuInAs122-NC12.999,EuInAs122-PhysRevLett.122.256402}.

Furthermore, the quantum anomalous Hall (QAH) insulator, characterized by dissipationless chiral edge states, exhibits quantized Hall conductivity in the absence of an applied magnetic field. It can be achieved in magnetic-doped TIs and quantum wells of magnetic Weyl semimetals (WSMs).
The magnetic WSM with a single pair of WPs, referred to as the `hydrogen atom' of WSMs, can exhibit large anomalous Hall conductivity and negative magnetoresistance driven by enhanced net Berry curvatures~\cite{CoSnS-NaturePhysics,CoSnS-NC,CoSnS-lzh,PhysRevB.102.161115,PhysRevX.11.031032,EuB6-PhysRevLett.124.076403}.
In the presence of inversion symmetry, the odd-valued $Z_4 = 1~(3)$ index stands for a semimetal phase. 
The theory of magnetic topological quantum chemistry has accelerated the discovery of magnetic topological materials~\cite{tqc2017,mtqc-Xu2020,mtqc2021,mtqc-gjc,PhysRevB.102.035144,PhysRevB.91.125139}. However, only the magnetic-doped Bi$_2$Se$_3$ family~\cite{Cr-Bi2Se3-PhysRevLett.109.266405,Cr-Bi2Se3-PhysRevLett.112.056801,Cr-Bi2Se3-acsnano.5b03980,Cr-Bi2Se3-acs.nanolett.6b03439} and few-layer MnBi$_2$Te$_4$ films~\cite{C3CE40643A,Gong_2019} have been fully verified by theory and experiment as intrinsic magnetic materials capable of realizing multiple topological phases. 

Recently, high-quality single crystals Eu$_3$In$_2$As$_4$ have been successfully synthesized, and magnetic measurements indicate that the material exhibits an antiferromagnetic (AFM) ground state, with a field-induced soft ferromagnetic (FM) state emerging above H = 1.5 T at 2 K~\cite{EuInAs324-yjy}. Moreover, the AFM transition is observed at a Néel temperature of 5 K.
The presence of Shubnikov–de Haas oscillations with a single frequency, along with a linear Hall resistance corresponding to the same carrier density, suggests that it belongs to the magnetic WSM phase, featuring only one pair of WPs near the Fermi level. 
In addition, magnetic Zintl Eu$_3$In$_2$As$_4$ nanowires have been successfully grown through the topotaxial mutual-exchange method~\cite{EuInAs324-nanowire}.
However, the magnetic ground state of Eu$_3$In$_2$As$_4$ remains uncertain, and a detailed analysis of its electronic structure is lacking. Therefore, further exploration is necessary to understand the potential magnetic topological phases and the interplay between magnetism and topology.

\begin{figure}
    \centering
     \includegraphics[width=0.85\linewidth]{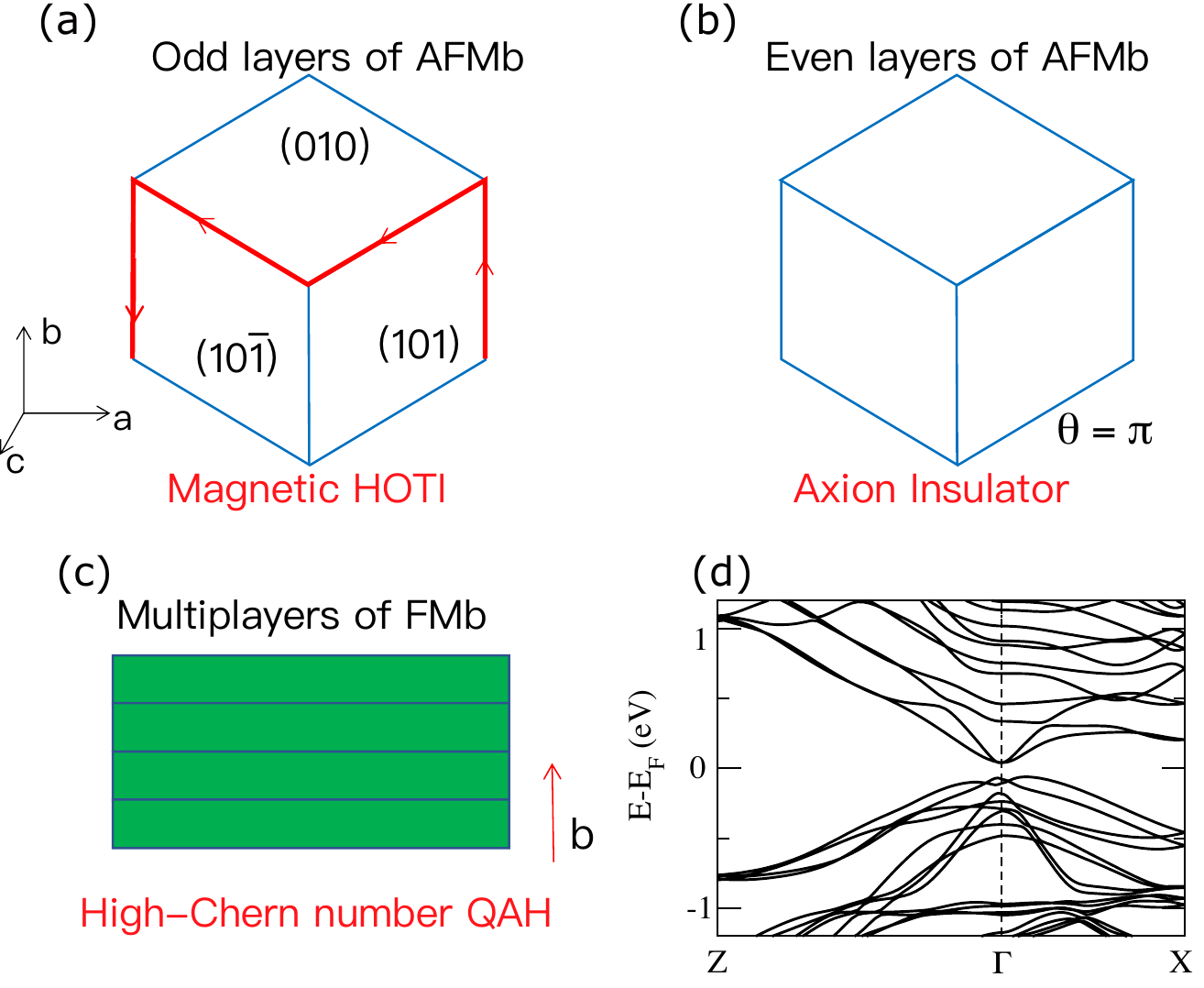}
    \caption{(Color online) Multiple magnetic topological phases. (a-c) Schematics for the magnetic HOTI, axion insulator, and high-Chern number QAH insulator for the topological AFMb ($Z_4=2$) and FMb ($Z_4=1$) states. To avoid the gapless surfaces states, the samples can be cut along the (101) and (10$\bar{1}$) surfaces of the AFMb Eu$_3$In$_2$As$_4$, whose surfaces can be fully gapped.
    The magnetic HOTI phase and axion insulator phase with the quantized axion angle ($\theta= \pi$) can be engineered in cubic samples with odd and even Eu$_3$-triplet layers, respectively. The high-Chern number QAH phase can be realized in quantum wells of the FMb state, such as multilayer films of Eu$_3$In$_2$As$_4$. (d) The nontrivial band structure of the FM Eu$_3$-triplet bilayer with the Chern number $C=1$.
    }
    \label{fig:1}
\end{figure}

In this letter, we predict through first-principles calculations that Eu$_3$In$_2$As$_4$ can exhibit multiple magnetic topological phases. It has an AFM ground state and is at the boundary of topological phase transitions. The magnetic higher-order topological insulator (HOTI), AI phase with the quantized axion angle ($\theta=\pi$), and high-Chern number QAH phases can be engineered in Eu$_3$In$_2$As$_4$ (Fig.~\ref{fig:1}). In the AFM state, it can be tuned to the AI phase with higher-order topology, characterized by the magnetic parity index $Z_4=2$. In the induced soft FM state, it becomes an ideal magnetic WSM with $Z_4=1$. We construct four-band $k \cdot p$ models to capture the underlying magnetic topological transitions.  In the end, the QAH effect is predicted to occur in the FM multilayer films of Eu$_3$In$_2$As$_4$.


%

\paragraph*{Methodology.---}
We performed the first-principles calculations based on the density functional theory (DFT) using the projector augmented wave (PAW) method~\cite{PAW1, PAW2}, implemented in the Vienna \emph{ab initio} simulation package (VASP)~\cite{VASP1, VASP2}.
The generalized gradient approximation with exchange-correlation functional of Perdew, Burke, and Ernzerhof~\cite{PBE} was adopted.
As the correlation effects of the localized $4f$ electrons of Eu should be considered, the DFT + U method was employed, with $U$ = 7 eV~\cite{PhysRevB.57.1505}.
The kinetic energy cutoff  was set to 500 eV for the plane wave basis.
The Brillouin zone (BZ) was sampled using the $\Gamma$-centered Monkhorst-Pack method~\cite{MPmethod} with an $8 \times 4 \times 10$ $\bk$-mesh in the self-consistent process.
The experimental lattice constants and atomic positions are used in the calculations~\cite{EuInAs324-yjy}. 
The irreducible representations (irreps) were obtained using the program \webirvsp\cite{GaojcIRVSP}, and the low-energy effective $k\cdot p$ models were constructed by \webvaspkp\cite{Zhang_2023}.

\begin{figure}
    \centering
     \includegraphics[width=0.98\linewidth]{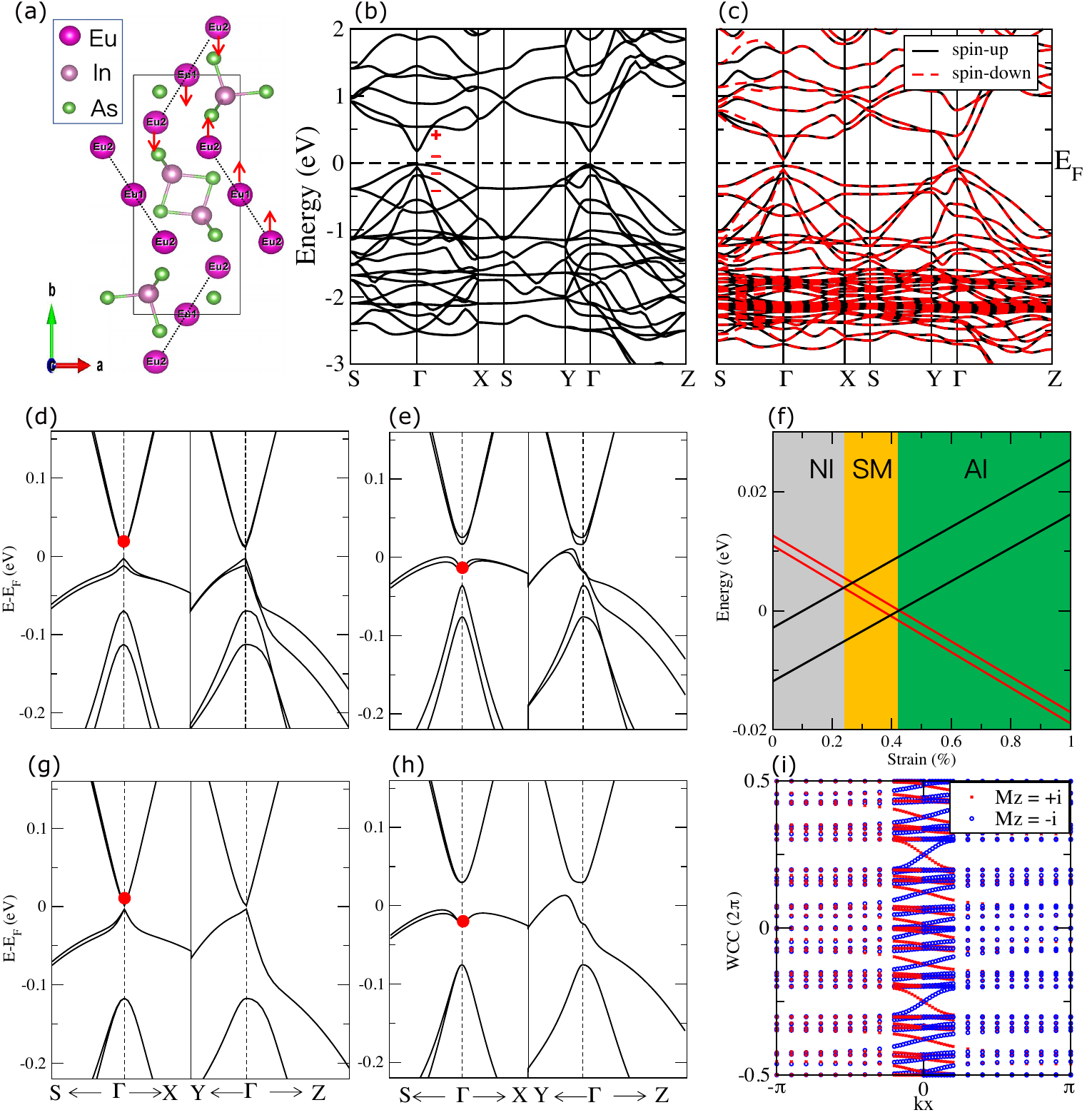}
    \caption{(Color online) Crystal structure and electronic band structures. (a) The crystal structure of Eu$_3$In$_2$As$_4$. The arrows indicate the magnetic directions of Eu atoms in the AFM configuration. Three Eu atoms form a Eu$_3$ triplet (dashed line), which can be treated as a layer in the $ac$ plane. The unit cell contains two Eu$_3$-triplet layers with opposite magnetic directions. (b,c) The NM and AFM band structures without spin-orbit coupling. The parities of the low-energy bands are depicted, indicating a topologically trivial phase. The AFMb and AFMc band structures (d,g) without and (e,h) with  1\% tensile strain in the presence of SOC. The even-parity bands are denoted by red dots at $\Gamma$. The 1\% tensile strain leads to a topological phase transition, from a normal insulator (NI, $Z_4=0$) to an axion insulator(AI, $Z_4=2$). (f) The diagram for the AFMb state under different tensile strains. A topological semimetal (SM) phase with $Z_4=1$ is found between the NI and AI phases. (i) The evolution of Wannier charge centers (WCCs) for the strained AFMc state, calculated using the 1D Wilson-loop method for the occupied bands with the mirror eigenvalue $M_z=+i$ $(-i)$ on the $k_z=0$ plane, which are plotted by blue circles (red crossings). The results show $C_{m=\pm i}=\mp1$ in the $k_z=0$ plane.
    }
    \label{fig:2}
\end{figure}

\begin{figure}
    \centering
     \includegraphics[width=0.85\linewidth]{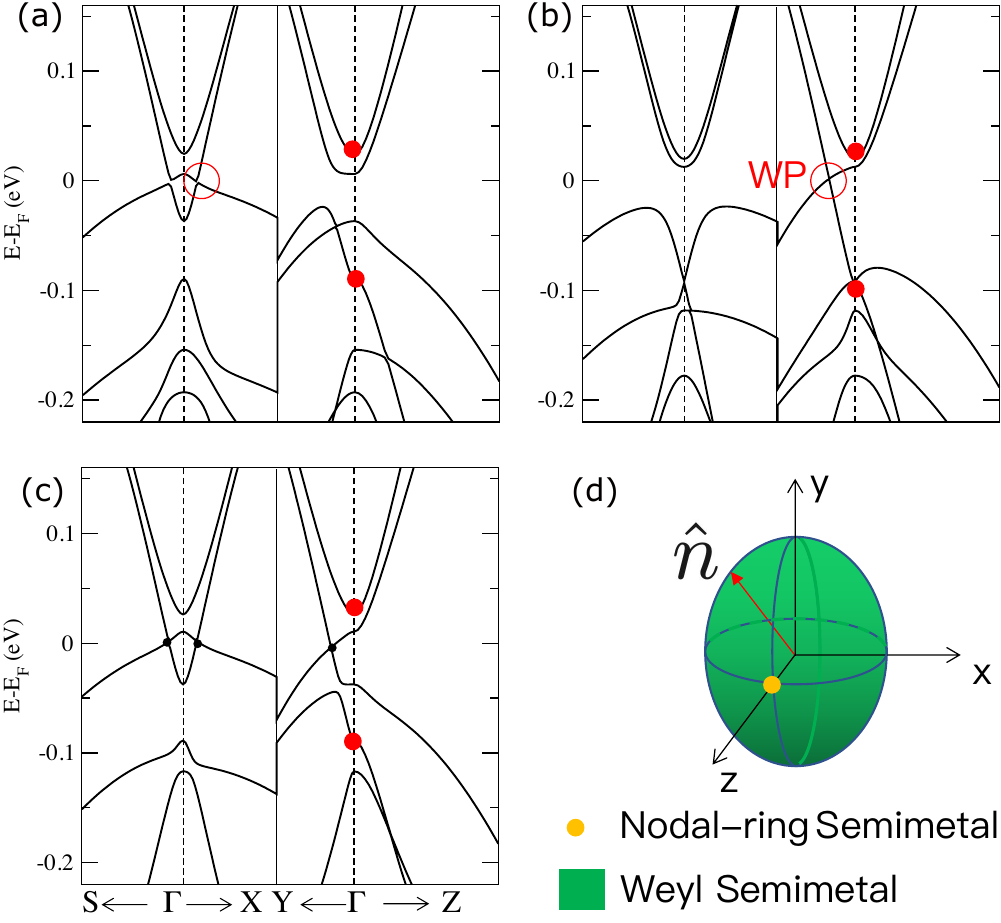}
    \caption{(Color online) FM band structures and topological phases. (a-c) The FMa, FMb, and FMc band structures. The even-parity bands are denoted by red dots at $\Gamma$. The parity index $Z_4=1$ indicates that they are magnetic topological semimetals. (d) The phase diagram for the FM state with a general spin direction $\hat n$. For $\hat n \parallel z$, it is a semimetal with a nodal ring; otherwise, it has a single pair of Weyl points.
    }
    \label{fig:3}
\end{figure}

\begin{figure}
    \centering
     \includegraphics[width=0.98\linewidth]{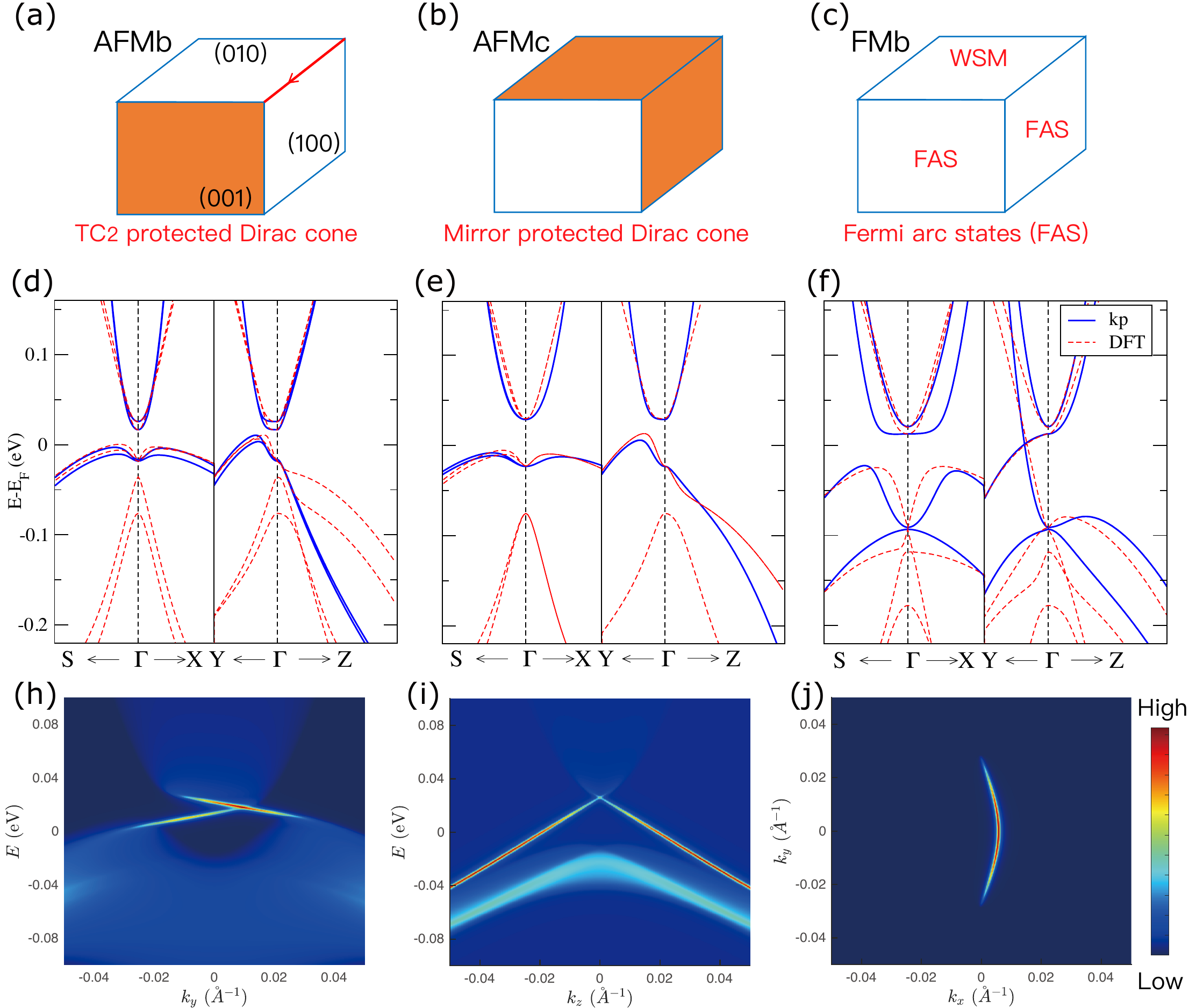}
    \caption{(Color online) Surface Dirac cones and Fermi arc states for the topological AFMb, AFMc ($Z_4=2$), and FMb ($Z_4=1$) states. (a-c) Schematics of surface gaps in the presence of SOC for each of them. The orange-colored surfaces denote gapless surface Dirac cone states. The red arrows denote the chiral hinge states. A $TC_2$-protected Dirac cone emerges on the (001) surface of the AFMb phase, while the $M_z$-protected Dirac cone emerges on the (100) and (010) surfaces of the AFMb phase. As the WPs of the FMb WSM phase are located along the $b$ direction, Fermi arc states are expected on the (100) and (001) surfaces. (d-f) The band dispersions of the $k\cdot p$ models, which fit the DFT band structures very well. The effective $k \cdot p$ models capture all the symmetries of the DFT bands and reproduce the topological phase transitions. (g-i) The surface spectra for the (001) surface of the AFMb state, and the (100) surface of the AFMc state, and the constant-energy contour for the (001) surface of the FMb state are computed. The topological surface Dirac cones and Fermi arc states are obtained.}
    \label{fig:4}
\end{figure}


\paragraph*{Crystal structure and band structures.---}
As illustrated in Fig.~\ref{fig:2}(a), the Zintl compound Eu$_3$In$_2$As$_4$ crystallizes in an orthorhombic lattice with the space group $Pnnm$ (SG 58), including an inversion $I$, a two-fold rotation $C_{2z}$, two screw rotations $\{C_{2x,2y}|\tau\}$, and two glide mirrors $\{IC_{2x,2y}|\tau\}$, where $\tau$ is the translation of one-half of a body diagonal $(\frac{a}{2},\frac{b}{2},\frac{c}{2})$. 
Especially, two Eu1 atoms are located at the $2c$ Wyckoff positions, and four Eu2 atoms are located at the $4g$ Wyckoff positions, as shown in Fig.~\ref{fig:2}(a). 
As suggested by the susceptibility measurements~\cite{EuInAs324-yjy}, the magnetic Eu$_3$In$_2$As$_4$ exhibits an AFM ground state at low temperatures, and Eu atoms exhibit the $\boldsymbol{S}=7 \mu _B$ magnetic moment, indicating the divalent Eu$^{2+}$ state. However, fitting a Curie–Weiss (CW) linear temperature dependence to the inverse susceptibility yields a positive Weiss constant $\theta_{CW}$, which is typical of FM correlations rather than AFM ones. This peculiarity is shared by several AFM Zintl compounds such as Eu$_3$In$_2$P$_4$, EuMn$_2$P$_2$, and EuSnP~\cite{Jiang2005,Payne2002,Payne2002_2}, and is commonly ascribed to short-range FM correlations. 
The bond length between the three Eu atoms (the Eu$_3$ triplet linked by a dashed line) is 3.83 \AA, while that between the triplets is 4.41 \AA, implying potentially different magnetic coupling. 
Therefore, the AFM configuration with parallel spin alignment within the triplet and anti-parallel spin alignment between the triplets is considered in our calculations. Yet, the real AFM order may be more complicated~\cite{PhysRevB.109.174404}, calling for further in-depth experimental investigations.


We first performed the nonmagnetic (NM) calculation by treating the seven Eu-$f$ electrons as core levels. The NM band structure is presented in Fig.~\ref{fig:2}(b). The parities of the low-energy bands are clearly denoted. The lowest conduction band has even parity, while the highest three valence bands have odd parity, consisting of In $p$ states. Based on the parity criterion, the weak and strong TI indices are computed as (000; 0), indicating a topologically trivial phase without band inversion. The band gap is 200 meV. After considering spin polarization, the trivial gap decreases to 92 meV in the AFM band structure of Fig.~\ref{fig:2}(c).
In the subsequent calculations with spin-orbit coupling (SOC), we find that, besides spin polarization, the SOC effect is crucial for topological phase transition, and further reduces the band gap.

\paragraph*{Altermagnet and symmetry.---}
In the AFM configuration of Fig.~\ref{fig:2}(a), the spin-up and spin-down atoms are related by $\{C_{2x/2y}|\tau\}$ or $\{IC_{2x/2y}|\tau\}$, rather than by either inversion or pure translation symmetry. 
In other words, $T\{IC_{2x/2y}|\tau\}$ (or $T\{C_{2x/2y}|\tau\}$) is respected, and the AFM state is an altermagnet.
Therefore, spin splitting occurs in the AFM band structure even without SOC, which is presented in Fig.~\ref{fig:2}(c).
We find that the spin-splitting bands are along S-$\Gamma$. But, the energy bands remain spin-degenerate along other high-symmetry lines in Fig.~\ref{fig:2}(c), because of the collinear spin-space group (SSG). The SSG operator $\{T||IC_{2x/2y}|\tau\}$ ensures that all the bands are spin-degenerate in $k_x=0,\pi$ and $k_y=0,\pi$ planes. Therefore, we conclude that this altermagnet exhibits tiny spin splitting.
We have computed the total energies of the AFM and FM states, the results show that the AFM state is more favorable, with the FM state being approximately 25 meV higher in energy per unit cell, which is consistent with the experimental data~\cite{EuInAs324-yjy}.

\paragraph*{Axion insulator under stains.---}

Once SOC is included, one can compare the energies for different spin directions, to check the magnetic anisotropy. They are nearly degenerate from our calculations with the energy difference being less than 1 meV. This is ascribed to the fact that the magnetic bands of Eu $4f$ orbitals are localized far from the Fermi level ($E_F$) in Fig.~\ref{fig:2}(c), making minimal influence on the low-energy bands at $E_F$. However, the direction of spin polarization is crucial for the magnetic symmetry, which in turn affects the gaps in the surface Dirac states.

In the AFM configuration with $z$-directed spin polarization in the presence of SOC, the magnetic system belongs to the magnetic space group (MSG) 58.1.471 (in OG convention~\cite{mtqc-gjc}). It belongs to a type-I MSG without time reversal symmetry ($T$), which contains all eight symmetries, such as $\{C_{2x,2y}|\tau\}$ and $I$. 
The energy bands at $\Gamma$ are doubly degenerate, due to the double point group of $D_{2h}$.
In the AFM with $y$-directed ($x$-directed) spin polarization, the system contains $TC_{2z}
$, $I$, $C_{2x}$ ($C_{2y}$) symmetries, corresponding to MSG 58.6.476. The bands at $\Gamma$ are non-degenerate, corresponding to the double point group $C_{2h}$. The antiunitary symmetry $TC_{2z}$ does not cause any degeneracy in the co-representations. 
In the AFM configuration with an arbitrary spin direction, the system still possesses $I$ symmetry. The magnetic parity index $Z_4$ works for all the cases.

The AFM band structures with $y$-directed (AFMb) and $z$-directed (AFMc) spin orientations, in the presence of SOC, are presented in Figs.~\ref{fig:2}(d,g), where the even-parity bands are depicted by red dots. The computed $Z_4=0$ indicates that these states are topologically trivial with a tiny band gap (about 3 meV), and two lowest conduction states are of even parity at $\Gamma$.
However, the topological phase transition can be easily tuned by applying strain. 
The corresponding band structures under strain are presented in Figs.~\ref{fig:2}(e,h). The two even-parity states become occupied, indicating the topological band inversion.
Our calculations show that under 1\% tensile strain, the system becomes topological, corresponding to the AI phase with $Z_4 = 2$, indicating the higher-order topology. Especially, in the $k_z=0, \pi$ planes of AFMb/AFMa state, the 2D Stiefel-Whitney (SW) insulator is well defined in the presence of both $TC_{2z}$ and $I$ symmetries. The 3D strong SW topological invariant ($Z_2$-type) is defined as $\Delta=w_2 (k_z=0)- w_2 (k_z=\pi)$, which is identical to $Z_4 \mod 2$ for the insulators~\cite{PhysRevB.92.081201,PhysRevLett.121.106403,Ahn_2019_CPB,PhysRevX.9.021013,QsWu_science_2019,Pan_Mingxiang_npjCM_2022,Guo_npjQM_2022,PhysRevB.91.161105,PhysRevB.99.235125}.
Therefore, the $Z_4=2$ phase in the AFMb/AFMa state corresponds to a 3D strong SW insulator with $\Delta=1$.

In the topological AFMb state, the (100) and (010) surfaces are gapped due to the absence of $M_z$ $(\equiv IC_{2z})$ symmetry. 
However, on the (001) surface, an unpinned Dirac point is protected by $TC_{2z}$ symmetry~\cite{fan_nc,PhysRevLett.119.246402,FrankSchindler_2018,ChenFang_2019}, consistent with the 3D strong SW insulator phase.  
In the presence of $I$ symmetry, the two (010)/(100) surfaces in Fig.~\ref{fig:4}(a) exhibit opposite masses. The chiral hinge state is expected on two of the four [001] hinges as the domain boundary. In the topological AFMc state, the mirror Chern numbers are calculated. The values $C_{M_z=\pm i}=\mp 1$ are obtained in Fig.~\ref{fig:2}(i) for $k_z = 0$, while $C_{M_z=\pm i}=0$ for $k_z=\pi$. The (100) and  (010) surfaces are gapless with a single Dirac cone on the $M_z$-preserved surface $k$-paths. However, the (001) surface is fully gapped, without any symmetry protected Dirac cones, as illustrated in Fig.~\ref{fig:4}(b).

\begin{table}[!htb]
\begin{ruledtabular}
    \centering
    \caption{The $k\cdot p$ parameters of strained $H_{\rm AFMb}$, strained $H_{\rm AFMc}$, and pristine $H_{\rm FMb}$.}
        \begin{tabular}{lrrr}
            Parameter& AFMb & AFMc &FMb \\
            \hline
            $b_1(eV)$& $0.0019$ & $0.0028$&$0.0203$\\
            $b_2(eV)$& $-0.0192$ & $-0.0260$&$0.0026$ \\
            $t_x(eV\cdot\text{\AA})$ & $2.0$ &$1.5$ &$0.8$\\
            $t_y(eV\cdot\text{\AA})$ & $0.7$ &$0.9$ &$0.5$\\
            $t_z(eV\cdot\text{\AA})$ & $3.0$ & $3.0$&$2.0$\\
            $A(eV\cdot\text{\AA})$ && $0.3$&\\
            $m_1^x(eV\cdot\text{\AA}^2)$ & $80.0$ &$33.0$ &$20.0$\\
            $m_1^y(eV\cdot\text{\AA}^2)$ & $45.0$ &$42.0$ &$50.0$\\
            $m_1^z(eV\cdot\text{\AA}^2)$ & $11.0$ &$42.0$ &$25.0$\\
            $m_2^x(eV\cdot\text{\AA}^2)$ & $83.0$ &$35.0$ &$25.0$\\
            $m_2^y(eV\cdot\text{\AA}^2)$ & $55.0$ &$51.0$ &$62.0$\\
            $m_2^z(eV\cdot\text{\AA}^2)$ & $13.5$ &$51.0$ &$30.0$\\
            $f_1(eV)$ & $-0.0019$ & &$0.0014$\\
            $f_2(eV)$ & $0.0027$ & & $0.0544$\\
        \end{tabular}
        \label{kp-parameters}
\end{ruledtabular}
\end{table}

\paragraph*{Magnetic Weyl semimetal with external fields.---}
The FM state can be induced by an external magnetic field. 
The band structures of FM states along the $x$ (FMa), $y$ (FMb) and $z$ (FMc) directions are presented in Figs.~\ref{fig:3}(a-c). The $I$ symmetry is always preserved. $Z_4 = 1$ is obtained, indicating a topological semimetal phase. In the FMb (FMa) state, a single pair of WPs is located along the $y$ ($x$) direction and protected by $\{C_{2y}|\tau\}$ ($\{C_{2x}|\tau\}$) symmetry, making it an ideal magnetic WSM.
The WPs are found at ($\pm$0.013 $\rm{\AA}^{-1}$, 0, 0) and (0, $\pm$0.026 $\rm{\AA}^{-1}$, 0) for the FMa and FMb states respectively, as enclosed by red circles in Figs.~\ref{fig:3}(a,b).
In the FMc state, the gapless points form a nodal ring in the $xy$ plane, protected by  $M_z$ symmetry. As part of the nodal ring, band crossings occur along S-$\Gamma$, $\Gamma$-X, and Y-$\Gamma$ lines, shown as black dots in Fig.~\ref{fig:3}(c). In the FM state with an arbitrary spin direction, the $Z_4=1$ guarantees the appearance of a single pair of WPs. The Fermi arc states are expected on the surfaces shown in Fig.~\ref{fig:4}(c).

\paragraph*{Effective $k\cdot p$ model and topological phase transition.---}
The low-energy bands in the NM state belong to the $\Gamma_1^+$ and $\Gamma_3^-$ irreps of the point group $D_{2h}$. Including the spin degree of freedom and spin-orbit coupling, the four-band effective model at $\Gamma$ for the magnetic states is obtained using the theory of invariants,
 \begin{equation}
 \begin{aligned}
    H_{\rm FM||\hat n}&= H_0+ H_1(\hat{\sigma}_{n}), \\
    H_{\rm AFMc}&= H_0+H_2, \\
    H_{\rm AFMb/a}&=H_0+H_1(\sigma_{x/y}),
\end{aligned}
\label{kpmsg}
\end{equation} 
where 
\begin{equation}
    \begin{aligned}
    H_0=&\left(b_1+\sum_{i=x,y,z}b_1^ik_i^2\right)\sigma_0\otimes\tau_0\\
    &+\left(b_2+\sum_{i=x,y,z}b_2^ik_i^2\right)\sigma_0\otimes\tau_3\\
    &+t_xk_x\sigma_y\otimes\tau_x+t_yk_y\sigma_x\otimes\tau_x+t_zk_z\sigma_0\otimes\tau_y,\\
    H_1(\hat{\sigma}_{n})&=f_1\hat{\sigma}_n\otimes\tau_0+f_2\hat{\sigma}_n\otimes\tau_3,\\
    H_2&=A(k_x\sigma_y\otimes\tau_y+k_y\sigma_x\otimes\tau_y+k_z\sigma_0\otimes\tau_x).
    \end{aligned}
    \label{kpmodel}
\end{equation}
The $\sigma_i$ and $\tau_i$ are Pauli matrices acting on the spin and orbital spaces. $\hat n=(n_x,n_y,n_z)$ is the magnetic direction vector (normalized) and $\hat{\sigma}_{n}=n_x\sigma_x+n_y\sigma_y+n_z\sigma_z$. By fitting the DFT band structures, the values of the $k\cdot p$ parameters for topological $H_{\rm AFMb}$, $H_{\rm AFMc}$, and $H_{\rm FM}$ are shown in Table~\ref{kp-parameters}. 
As reported in Ref.~\cite{EuInAs324-yjy}, the AFMb state is compatible with the FMa state, which forms the canted FM state of MSG 58.6.476. That is why the $H_{\rm AFMb}$  and $H_{\rm FMa}$ models share the same form.
The strain effect of the AFMb state can be parametrized by $b_2$. As $b_2$ changes from positive to negative, the band inversion occurs smoothly, from the NI ($Z_4=0$) to topological semimetal ($Z_4=1$) and then to AI ($Z_4=2$), as shown in Fig.~\ref{fig:2}(f).
In the FM state with a general spin direction $\hat n$, the topological semimetal is fully captured by the $k\cdot p$ model in Eq.~(\ref{kpmsg}), yielding a nodal-ring semimetal phase in the $z$ direction and the WSM phase in other directions, such as the $a$ and $b$ directions.

\paragraph*{Quantum anomalous Hall effect in quantum wells.---}

By treating the Eu$_3$ triplet as a layer in the $ac$ plane, the nature stacking of the topological AFMb state in the $b$ direction can give rise to the magnetic HOTI and AI.
In the topological AFMb state with $Z_4=2$, gapped states are observed on the (010) and (100) surfaces, while a gapless $TC_2$-protected Dirac cone appears on the (001) surface, as shown in Fig.~\ref{fig:4}(a). To avoid gapless surface states, the samples can be cut along the (101) and (10$\bar{1}$) surfaces, as shown in Fig.~\ref{fig:1}(a). Thus, all surfaces
are gapped. One can then generate the magnetic HOTI (odd layers) and AI (even layers) in these geometrical samples with varying numbers of Eu$_3$-triplet layers in the $b$ direction.
When there is an odd number of layers, the top and bottom (010) surfaces have the same $\sigma=+ \frac{1}{2} e^2/h$
Hall conductance, corresponding to a $C=1$ Chern insulator. 
Due to the presence of $I$ symmetry, the chiral hinge state is expected, as shown in Fig.~\ref{fig:1}(a), which conducts the quantized Hall conductance.
When there is an even number of layers, the top and bottom (010) surfaces have the opposite Hall conductance, $\sigma=\pm \frac{1}{2} e^2/h$.
Meanwhile, the $I$ symmetry is not respected any more. At this point, all the surfaces and hinges can be gapped. Instead, the quantized chiral axion $\theta=\pi$ can be detected.

As a magnetic WSM in the FMb state of Eu$_3$In$_2$As$_4$, the anomalous Hall effect is an exotic consequence. The magnitude of anomalous Hall conductivity is estimated by $\sigma_{xz} = (e^2/h)(\Delta k^W_y/2\pi)$, where $\Delta k^W_y$ is the distance between the WPs in the $y$ direction. Additionally, the 3D WSM is reviewed as a stack of two-dimensional Chern insulators with strong coupling in the stacking direction. The band structure of the Eu$_3$-triplet bilayer is shown in Fig.~\ref{fig:1}(d), revealing a Chern insulator with a 101 meV nontrivial gap. The Chern number $C = 1$ is computed using the 1D Wilson-loop method. Considering the confinement effect in quantum wells, the high-Chern number QAH effect can be achieved by increasing the thickness of the quantum wells ~\cite{EuB6-PhysRevLett.124.076403,EuB6-PhysRevX.11.021016}. 
Thus, we propose that the high-Chern number QAH effect can be generated in multilayer films of soft FM Eu$_3$In$_2$As$_4$ on an insulating magnetic substrate.

\paragraph*{Conclusion.---}
In conclusion, we predict that magnetic Eu$_3$In$_2$As$_4$ lies at the boundary between different topological phases.
It is a trivial AFM semiconductor with a small gap of approximately 3 meV. Strain and magnetic fields can drive it into multiple distinct magnetic topological phases, such as the AI with $\theta=\pi$, HOTI with $Z_4=2$ and an ideal magnetic WSM.  
Strain is typically induced by the experimental growth conditions and substrate choice.
Furthermore, the soft FM state can be induced by an insulating magnetic substrate. Due to limitations in the accuracy of magnetic DFT+U calculations and the magnetic configurations considered, the exact $E_g$ is difficult to predict. However, multiple topological phases are accessible under modest experimental conditions.

\ \\
\noindent \textbf{Acknowledgments}
This work was supported by 
the National Key R\&D Program of China (Grants No. 2022YFA1403800, 2021YFA1400401),
the National Natural Science Foundation of China (Grants No. 12188101, U22A6005, U2032204), the Strategic Priority Research Program of Chinese Academy of Sciences (Grant No. XDB33000000), 
the Center for Materials Genome, and the Synergetic Extreme Condition User Facility (SECUF).

\noindent \textbf{Note added}
In the preparation of this work, we noticed another similar work on Eu$_3$In$_2$As$_4$~\cite{yan}, where the antiparallel spin alignment within the triplet is considered. Although the band gaps change a little bit, the symmetry analysis and topological phase transitions remain the same.

%

\end{document}